\documentclass[structabstract]{aa}
\usepackage{graphicx}
\usepackage{txfonts}
\begin{document}
   \title{Edge-on disk around the T Tauri star [MR81] H$\alpha$ 17 NE in 
          CrA \thanks{Based on observations collected at the 
European Southern Observatory, Chile in runs 
67.C-0213, 71.C-0206(D), 71.C-0522(B), 073.C-0164(A), 
073.C-0167(B) and 081.C-0393(A).}}

\author{R. Neuh\"auser \inst{1}, 
        S. Kr\"amer \inst{1},
        M. Mugrauer \inst{1},
        R. K\"ohler \inst{2,3},
        T.O.B. Schmidt \inst{1},
        M. Ammler-von Eiff \inst{4},
        J. Alves \inst{5},
        S. Fiedler \inst{1},
 \and N. Vogt \inst{6,7} }

\institute{Astrophysikalisches Institut und Universit\"ats-Sternwarte, 
Friedrich-Schiller-Universit\"at Jena, Schillerg\"asschen 2-3, 07745 Jena, Germany \\
     \email{rne@astro.uni-jena.de}
\and
ZAH Landessternwarte, K\"onigstuhl, 69117 Heidelberg, Germany
\and
Max-Planck-Institut f\"ur Astronomie, K\"onigstuhl 17, 69117 Heidelberg, Germany
\and
Centro de Astronomia e Astrof\'{\i}sica da Universidade de Lisboa, Tapada da Ajuda, 1349-018 Lisboa, Portugal
\and
Calar Alto Observatory, Centro Astronomico Hispano-Aleman, C/q Jesus Durban Remon 2-2, 04004 Almeria, Spain
\and
Departamento de F\'isica y Astronom\'ia, Universidad de Valpara\'iso,
Avenida Gran Breta\~na 1111, Valpara\'iso, Chile
\and
Instituto de Astronom\'ia, Universidad Catolica del Norte, Avda. Angamos 0610,
Antofagasta, Chile
}

\date{Received 1 Aug 2008; accepted 17 Nov 2008}

\abstract
{}
{Using the speckle camera SHARP at the 3.5m ESO NTT, 
K\"ohler and collaborators 
found an object $\sim 3.5$ mag fainter in K only $1.3^{\prime \prime}$ north-east 
of the T Tauri star [MR81] H$\alpha 17$ in the Corona Australis
(CrA) star-forming region, which could be either a brown dwarf or a T Tauri star 
with an edge-on disk. We attempt to study this faint object in detail.}
{We acquired deep VLT NACO near-infrared images at three epochs to determine,
whether [MR81] H$\alpha 17$ and the nearby faint object are comoving and to
measure the infrared colors of both objects. We obtained optical and infrared
spectra of both objects with the VLT using FORS and ISAAC, respectively,
to determine spectral types and temperatures as well as ages and masses.}
{The T Tauri star [MR81] H$\alpha 17$ and the faint nearby object have a 
projected separation of 1369.58 mas, i.e. 178 AU at 130 pc. 
They share the same proper motion ($\sim 5 \sigma$), 
so that they most certainly form a bound binary pair.
The apparently fainter component [MR81] H$\alpha 17$ NE has a spectral type 
of M2e, while the apparently brighter component [MR81] H$\alpha 17$ SW, 
the previously known T Tauri star, has a spectral type of M4-5e.
We can identify a nearly edge-on disk around [MR81] H$\alpha 17$ NE by visual 
inspection, which has a diameter of at least 30 to 50 AU.
We are able to detect strong emission lines in [MR81] H$\alpha 17$ NE,
which are almost certainly due to ongoing accretion. The NE object is 
detectable only by means of its scattered light.
}
{If both objects are co-eval (2-3 Myr) and located at the same 
distance ($\sim 130$ pc as CrA), then the apparently fainter 
[MR81] H$\alpha 17$ NE is more massive (primary) component
with a nearly edge-on disk and the apparently brigther component 
[MR81] H$\alpha 17$ SW is less massive (companion). 
Both are low-mass T Tauri stars with masses of $\sim 0.5$ 
and $0.23 \pm 0.05$~M$_{\odot}$, respectively.}

\authorrunning{Neuh\"auser et al.}


\maketitle

%

\section{Introduction: [MR81] H$\alpha$ 17 in CrA}

The star-forming region Corona Australis (R CrA or CrA) harbors
dozens of young intermediate- to low-mass stars (and brown dwarfs) from spectral types 
B8 down to M8.5. They have an age of between one and a few Myrs and at a distance 
of appoximately 130 pc (see Neuh\"auser \& Forbrich (2008) for a recent review). 

The first few low-mass, pre-main sequence stars, so-called T Tauri stars
(TTS) in this star-forming region were found by early H$\alpha$ and
infrared (IR) imaging surveys by Knacke et al. (1973), Glass \& Penston (1975),
and Marraco \& Rydgren (1981). The latter found an emission-line object, 
nowadays called [MR81] H$\alpha$ 17 (MR81 for Marraco \& Rydgren 1981,
sometimes also [MR81] HA 17).
It is located at $\alpha = 19^{\rm h} 10^{\rm m} 43.4^{\rm s}$ and 
$\delta = -36^{\circ} 59^{\prime} 09^{\prime \prime}$ for J2000.0.
The catalogs GSC, USNO, NOMAD, DENIS, and 2MASS
provide BVRIJHK magnitudes for the unresolved binary.

Patten (1998) determined the spectral type M3-5 for the unresolved binary.
The USNO B1.0 and NOMAD catalogs provide a proper motion of
$2.0 \pm 25.0$ mas/yr in right ascension and $-44.0 \pm 13.0$ mas/yr in declination,
while Ducourant et al. (2005) indicate a proper motion of
$10 \pm 12$ mas/yr in right ascension and 
$-24 \pm 12$ mas/yr in declination.
We decided to use the weighted mean, 
i.e. $8.5 \pm 10.8$ mas/yr in right ascension
and $-33.2 \pm 8.8$ mas/yr in declination,
which is consistent with kinematic membership to CrA
(see e.g. Neuh\"auser et al. 2000).
 
\begin{table*}
\caption[]{Observations log}
\begin{tabular}{lrlrrrlccc} \hline
Telescope/ & observing     & filter and & DIT & NDIT & NINT & FWHM  & pixel scale & detector orientation & remarks \\ 
instrument & date          & obs. mode  & [s] &      &      & [mas] & [mas/px]  & [$^{\circ}$] (a) & \\ \hline
NTT/SHARP  & 6-July-2001  & K imaging  & 0.5 & 250  & 2    & 671  & $49.5 \pm 0.2$ & $90.2 \pm 0.2$ & (b) \\
VLT/ISAAC  & 27-July-2002 & J imaging  & 1.8 & 33   & 7    & 652  & $147.856 \pm 0.047$ & $179.96 \pm 0.01$ &(c) \\
VLT/ISAAC  & 27-July-2002 & H imaging  & 1.8 & 33   & 7    & 801  & $147.856 \pm 0.047$ & $179.96 \pm 0.01$ &(c) \\ 
VLT/ISAAC  & 27-July-2002 & K imaging  & 1.8 & 33   & 7    & 717  & $147.856 \pm 0.047$ & $179.96 \pm 0.01$ &(c) \\
VLT/FORS1  & 27-Mar-2003  & red spectrum & 790 & (d)  & 3  & -    & 200 & 0 & (e) \\
VLT/NACO   & 4-Sept-2003  & K imaging  & 1.7 & 40   & 10   &  83  & $13.24 \pm 0.02$ & $-0.02 \pm 0.09$ & (g) \\
VLT/ISAAC  & 5-May-2004   & K spectrum & 0.11 & 9   & 26   & -    & 71 & 0 & (e) \\
VLT/NACO   & 25-June-2004 & K imaging  & 3.5 & 20   & 27   & 116 & $13.23 \pm 0.05$ & $0.14 \pm 0.25$ & (f) \\
VLT/NACO   & 13-June-2008 & J imaging  & 20 & 3   & 22   & 391 & $13.243 \pm 0.056$ & $0.73 \pm 0.40$ & (g) \\ 
VLT/NACO   & 13-June-2008 & H imaging  & 15 & 4   & 30   & 285 & $13.243 \pm 0.056$ & $0.73 \pm 0.40$ & (g) \\
VLT/NACO   & 13-June-2008 & K imaging  & 10 & 6   & 30   & 231 & $13.243 \pm 0.056$ & $0.73 \pm 0.40$&(g)\\ \hline
\end{tabular}

Remarks: Total exposure time are DIT (individual Detector Integration Times) times
NDIT (number of DITs, together saved in one fits file) times NINT (number of jitter positions and files).
(a) Measured from North over East to South. Positive detector orientations mean
that they are to be added to values measured on raw frames.
(b) Speckle imaging reported in K\"ohler et al. (2008),
$500 \times 0.5$s on two detector positions each. For astrometric calibration, see footnote 1.
(c) Astrometric calibration with 2MASS sources in the field-of-view.
(d) Three spectra of 790s exposure each.
(e) Nominal pixel scale and detector orientation from the fits header.
(f) Astrometric calibration from Neuh\"auser et al. (2005).
(g) Astrometry done as in Neuh\"auser et al. (2008) with Galactic Center 
images for 2003 and images of the Hipparcos binary HIP 73357 for 2008.
\end{table*}

Using the speckle camera SHARP at the ESO 3.5m NTT on 6 July 2001, 
K\"ohler et al. (2008) detected a faint object close to [MR81] H$\alpha$ 17. 
The apparently fainter object is $\sim 3.5$ mag fainter in K and
located at a separation of $\sim 1.3 ^{\prime \prime}$ NE from the apparently 
brighter object [MR81] H$\alpha$ 17, which we now label
[MR81] H$\alpha$ 17 SW; the apparently fainter object is
now refered to as [MR81] H$\alpha$ 17 NE.
Figure 3 of K\"ohler et al. (2008) displays their SHARP K-band image
of [MR81] H$\alpha$ 17 NE \& SW, which has lower angular resolution
and a lower signal-to-noise (S/N) ratio than our new VLT AO image (Fig. 1).

The apparently fainter object can be a young star with (nearly) 
edge-on disk, a brown dwarf companion, or an unrelated background star, 
as speculated by K\"ohler et al. (2008).
We identify the fainter object as an early M-type T Tauri star
with a nearly edge-on disk and strong emission lines due to ongoing accretion.
We present the observations and data reduction in Sect. 2 and the results in Sect. 3.
 
\section{Observations and Data Reduction}

The apparently fainter object next to [MR81] H$\alpha$ 17 was discovered on
the K-band speckle images taken with SHARP-I 
(System for High Angular Resolution Pictures number I)
of the Max-Planck-Institut for Extraterrestrial Physics (Hofmann et al. 1992)
at the 3.5m New Technology Telescope (NTT)
of the European Southern Observatory (ESO) on La Silla, Chile
(see K\"ohler et al. (2008) for details on the observations,
data reduction, and the resulting image).\footnote{The SHARP data
were reduced independently by K\"ohler et al. (2008) with the speckle
interferometry technique and by Kr\"amer (2008) with a simple shift+add
technique; they determined and used two slightly different 
pixel scales and detector orientations (Kr\"amer uses 
$50.44 \pm 0.64$ mas/pixel for right ascension,
$48.02 \pm 0.64$ mas/pixel for declination,
and a detector orientation of $90.6 \pm 0.5 ^{\circ}$ obtained by
T. Ott (priv. comm.) from the Galactic Center images of the same night;
see Table 1 for the astrometric solution of K\"ohler et al. (2008); 
with slightly different calibrations, 
the results on separation and PA also differ slightly; 
in this paper, we use the mean of their results for [MR81] H$\alpha$ 17,
listed in Table 2.}
To find out whether the apparently fainter object could be either a
sub-stellar companion or a star with an edge-on disk, follow-up
observations were necessary.

Follow-up imaging in JHK was completed one year later with
the Infrared Array Camera and Spectrograph (ISAAC) of
the ESO Very Large Telescope (VLT) on Cerro Paranal, Chile.
Images of higher angular resolution and higher S/N in JHK
were taken with the Adaptive Optics imager NACO (for NAOS CONICA
for Nasmyth Adaptive Optics System, NAOS, with 
COude NearInfrared Camera and Array, CONICA;
Rousset et al. 2003) at the ESO VLT in 2003, 2004, and 2008,
with the S13 camera, i.e. a $14^{\prime \prime} \times 14^{\prime \prime}$ field of view.
Optical and near-infrared spectra of the apparently fainter object
were obtained with the ESO VLT instruments 
FORS1 (FOcal Reductor and Spectrograph number 1)
and ISAAC, respectively (see table 1 for the entire observations log).

All science and flat field frames acquired were dark and/or bias corrected,
then the science frames were divided by a normalized flat field.
A shift+add procedure was applied to subtract the background and
to add up all frames in each filter for each observing run.
The same procedure was performed for spectroscopic (telluric)
and imaging (photometric and astrometric) standard stars.
Spectra were wavelength calibrated with lamp arc exposures,
divided by the telluric standard, and multiplied by a corresponding
spectral template.
For astrometric calibration of the imaging observations, we used 
astrometric standards observed in the same night, such as 
stars close to the Galactic Center or well-known Hipparcos binaries.
Resulting pixel scales and detector orientations are given in Table 1;
the errors include the effects of Gaussian centering errors in the science targets
and the astrometric standards as well as possible motion in the standards
(see e.g. Neuh\"auser et al. (2008) for details of typical astrometric 
calibration procedure). For data reduction, we used ESO eclipse and MIDAS 
for imaging data, and IRAF for spectra.

\section{Results}

We present and discuss the results for our
imaging (photometry and astrometry) data and
spectroscopy (physical parameters) data.

\begin{figure}
\includegraphics[angle=0,width=1\hsize]{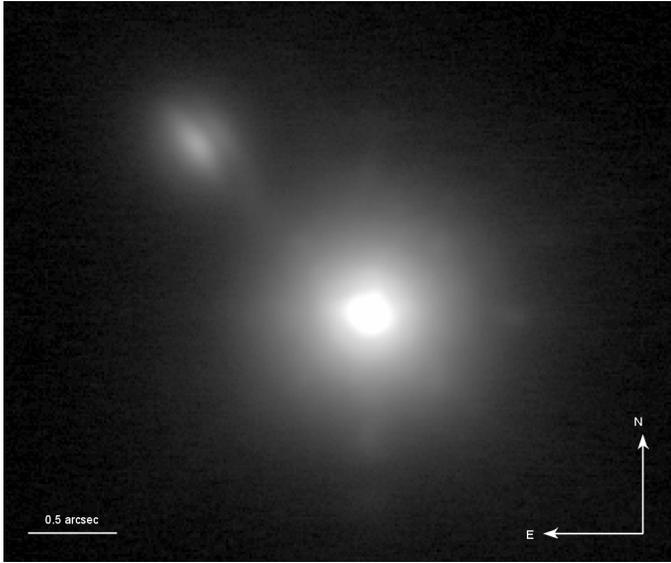}
\caption{VLT NACO K-band image of [MR81] H$\alpha$ 17 SW
(brighter object, K=9.60 mag) and NE (top left, K=13.24 mag) 
obtained in 
June 
2004 with the nearly edge-on disk. 
This disk is seen as dark lane extended by about 200 to 450 
mas (26 to 57 AU at 130 pc) along the NE-SW direction.
The 
dark lane thickness
is $\sim 100$ mas, i.e. 15 AU (possibly flared disk).
The SE surface of the disk reflects far more star light than
the NW surface; hence, the disk is not exactly edge-on,
but inclined by about $20^{\circ}$ from edge-on.
Angular resolution and S/N are insufficient for 
determining the disk properties accurately.
The faint dark lane of the edge-on disk can be seen better on 
a computer screen when using the original fits
file (www.astro.uni-jena.de/Users/rne/cra-disk).
See Fig. 2 for a 3D contour plot.
See Fig. 13 in the online appendix for a color-composite JHK image.}
\end{figure}

\begin{figure}
\includegraphics[angle=0,width=1\hsize]{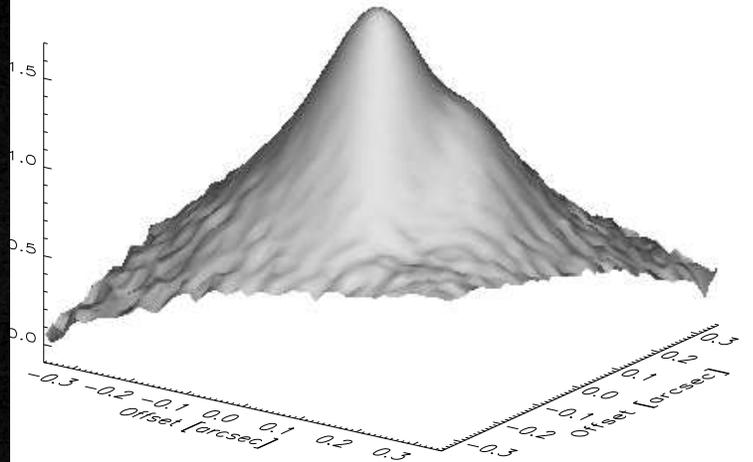}
\caption{Three-dimensional contour plot (log stretch) of the 
K-band brightness of [MR81] H$\alpha$ 17 NE obtained from Fig. 1.
The nearly edge-on disk, hardly visible as a dark lane in Fig. 1,
is visible here as a step. A higher resolution image may be taken
in the future with new AO instruments or larger telescopes.}
\end{figure}

\subsection{Imaging results: Photometry and astrometry}

In our NACO images (Fig. 1), the ellipsoidal shape of
the NE component is visible with a faint dark lane across the object. 
This image clearly resembles the typical appearence of 
a classical T Tauri star with (nearly) edge-on disk.

The first young stellar object displaying this pattern, HH 30 IRS in Taurus, 
was imaged by the Hubble Space Telescope (Burrows et al. 1996). Edge-on disks 
are also observable with ground-based near-IR imaging observations 
both with AO,
e.g. HK Tau/c (Stapelfeldt et al. 1998, Koresko 1998),
HV Tau C (Monin \& Bouvier 2000),
and LkH$\alpha$ 263 C (Jayawardhana et al. 2002) 
and seeing-limited, e.g.
2MASSI J1628137−243139 (Grosso et al. 2003).
Hence, this object is most probably a young star with a nearly edge-on disk.
The faint red object also resembles the appearance of so-called {\em infrared companions},
see e.g. Koresko \& Leinert (2001), a typical example of which 
is T Tau Sa (Koresko et al. 2000).

\begin{figure}
\includegraphics[angle=0,width=1\hsize]{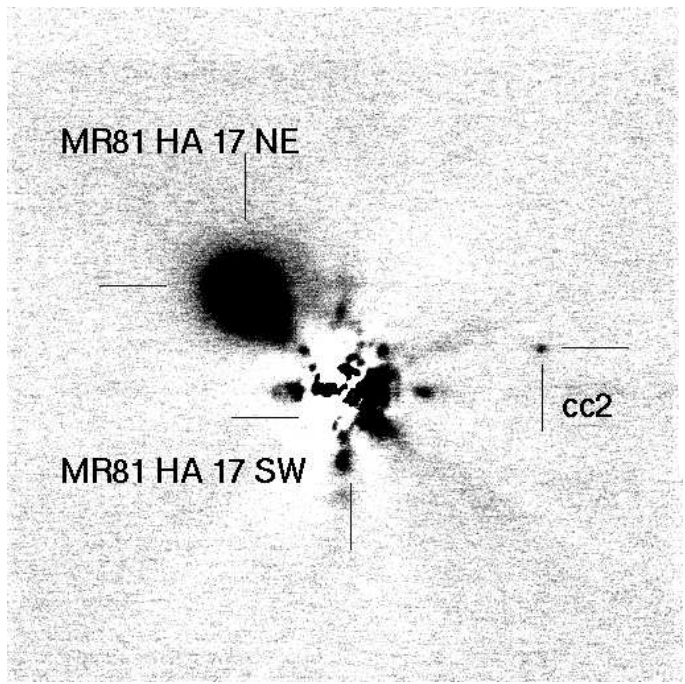}
\caption{VLT NACO K-band image of [MR81] H$\alpha$ 17
(obtained in June 2004)
after subtraction of the PSF of the brighter object (SW),
located here in the center of the image, 
the remaining speckles are seen as noise.
One can see the fainter object NE (top left, K=13.24 mag,
edge-on disk not seen here due to cuts selected to
see cc2) and also another much fainter object towards to WNW
(to the right, K=18.56 mag), called cc2 here. 
This is a companion candidate, which would
have only 1-2 Jupiter masses, if bound. However, we have
only one epoch available for this faint object,
because it is not detected in the other images.}
\end{figure}

Astrometric and photometric data for NE are obtained after
subtraction of the Point Spread Function (PSF) of the brighter star SW.
Our astrometric results for the objects [MR81] H$\alpha$ 17 SW \& NE 
are listed in Table 2. The brighter object (SW) has a FWHM of $116 \pm 4$ mas.
The position angle (PA) of the disk major axis is $47.0 \pm 0.2 ^{\circ}$,
where the extension along the disk lane axis is maximal, namely 247 mas,
while the extension perpendicular to the disk lane is only 181 mas
(both measured in the June 2004 NACO image);
this disk angle is closely aligned (within $2^{\circ}$) with the PA between the 
SW and NE sources. However, as seen in Fig. 4 \& 5, a more pole-on orbit 
of SW around NE is more likely than a more edge-on orbital inclination, 
so that disk and binary orbit were not coplanar.

The NACO image after subtraction of the PSF of SW is shown in Fig. 3.
We detect an additional faint object $\sim 2^{\prime \prime}$
NW of [MR81] H$\alpha$ 17 SW. This object is detected only in our
highest quality image, the NACO image of June 2004, not in any other image.
This faint object close to a bright star, for which only one epoch 
of data is available,
must be regarded as a companion candidate (cc); we call it [MR81] H$\alpha$ 17 SW/cc2
(cc1 was what is now [MR81] H$\alpha$ 17 NE).
This new object is $1988.714 \pm 6.97$ mas west of SW
and $420.1098 \pm 7.80$ north of SW,
corresponding to $\sim 264$ AU projected separation at 130 pc.
This separation is close to the separation between the stars
[MR81] H$\alpha$ 17 SW and NE; such an non-hierarchical triple
would probably be unstable.
With K=$18.56 \pm 0.1$ mag (and using B.C.$_{\rm K}=3.3$ mag as for
substellar L- and T-type objects), it would have a bolometric
luminosity (at 130 pc) of $\log L_{\rm bol}/L_{\odot} = -4.7 \pm 0.1$
and, hence, only about 1 to 2 Jupiter masses (according to either
Burrows et al. 1997 or Baraffe et al. 1998), for 2-3 Myrs of age.
However, not only are these models with evolutionary tracks
uncertain at such very young ages, we also should note that this
object is more likely to be a background object then another bound companion.
A 2nd epoch image should be taken to check for common proper motion.
In addition to NE and cc2, there are no other objects detected within
$5^{\prime \prime}$ around SW down to K=19 mag; faint objects within
$0.5^{\prime \prime}$ of SW cannot be excluded.

\begin{table}
\caption[]{Astrometric results}
\begin{tabular}{lccc} \hline
Date         & separation    & PA & Remark \\
             & [mas]         & [$^{\circ}$] & \\ \hline
6-July-2001  & $1313 \pm 32$ & $45.15 \pm 2.0$ & (a) \\ 
27-July-2002 & $1384 \pm 39$ & $45    \pm 1.9$  & (b) \\
4-Sept-2003  & $1369.5 \pm 7.3$ & $45.07 \pm 0.32$ & (b) \\
25-June-2004 & $1365.8 \pm 3.9$ & $45.4   \pm 0.25$  & (b) \\
13-June-2008 & $1373.4 \pm 7.2$ & $45.88 \pm 0.40$  & (b) \\ \hline
\end{tabular}

Remarks: (a) Means from data in Kr\"amer (2008) and 
K\"ohler et al. (2008) using slightly different SHARP pixel scales. (b) this work.
\end{table}

In Figs. 4 \& 5, we show the evolution in the separation and PA with time
to investigate whether the NE and SW components
form a common proper motion pair.
For the epoch 2002 (ISAAC), we use only the K-band data,
because the fainter object is hardly detected in J and H.

To test the background and companion hypotheses, 
we estimate the maximum possible orbital motion
for edge-on orbit (separation change) and pole-on orbit (PA change).
We assume a circular orbit. The most reliably measured separation (2004) is
$1365.8 \pm 3.9$ mas, i.e. $\sim 178$ AU at 130 pc,
yielding a $\sim 3000$ yr long orbit (for 0.6~M$_{\odot}$ total mass).
When using all data available, common proper motion appears possible;
only the SHARP data point regarding separation appears to be deviant by 1 to 2 $\sigma$.
When using only the NACO data, which provide the most accurate measurements,
the smallest errors, and are all taken with the same instrument
and same set-up (high angular resolution AO), 
both objects clearly exhibit common proper motion,
i.e. are most certainly bound and orbiting each other.

\begin{figure}
\includegraphics[angle=270,width=8.5cm]{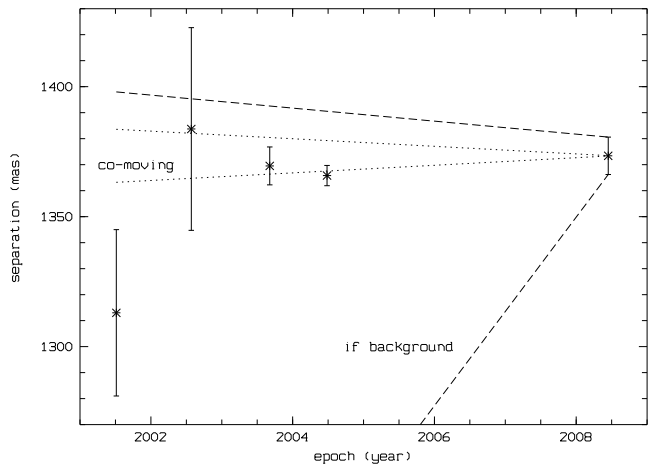}
\caption{Separation (in mas) versus observing epoch (in years) for
data listed in table 2. 
The dotted lines (starting from the 2008 data point opening to the
past) indicate maximum possible separation change due to orbital motion
in case of edge-on orbit. The dashed lines are for the background
hypothesis, if the brighter object (SW) would have moved according to
its (poorly) known proper motion, while the fainter object (NE)
would be a non-moving object;
we use the weighted mean of the USNO and NOMAD values for the proper 
motion of the SW object, namely $mu _{\alpha} = 8.5 \pm 10.8$ mas/yr 
and $mu _{\delta} = -33.2 \pm 8.8$ mas/yr;
the two dashed lines also start in 2008 and open to the past, 
they take into account the error in proper motion. 
The last three data points from NACO with smallest 
error bars are consistent with (nearly) constant separation,
the average being 1369.58 mas.}
\end{figure}

\begin{figure}
\includegraphics[angle=270,width=1\hsize]{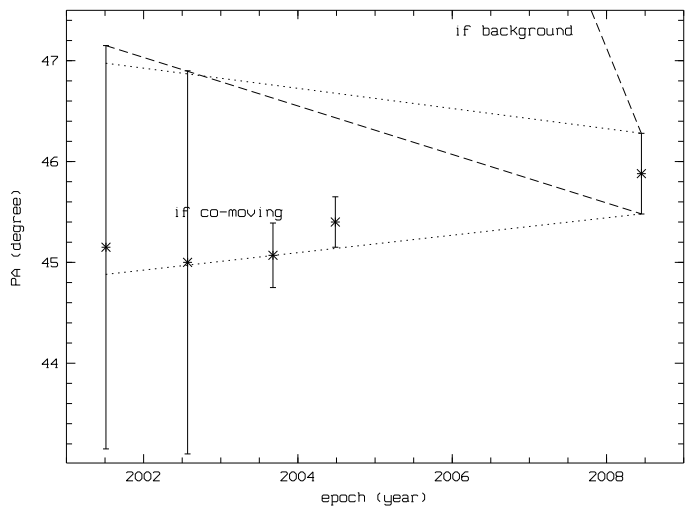}
\caption{Position angle (in degrees) versus observing epoch (in years) for
data listed in Table 2. The dotted lines (starting from the 
2008 datapoint and opening to the past) indicate maximum PA change due to 
orbital motion in the case of a pole-on orbit. The dashed lines indicate the background
hypothesis, if the brighter object (SW) has moved according to
its (poorly) known proper motion, while the fainter object (NE)
would be a non-moving object;
again, we use the weighted mean of the USNO and NOMAD proper motions for the SW object.
The datapoints from 2003 and 2004 are inconsistent with
the background hypothesis by $\sim 5 \sigma$ (together),
so that in this case we can reject the background hypothesis.
The datapoints are fully consistent with common proper motion.
}
\end{figure}

Photometric results are as follows.
By comparing aperture photometry for the apparently brighter object SW and
the apparently fainter object NE (after subtraction of the PSF
of SW), we derive magnitude differences as follows:
$\Delta$ J $\simeq 5.01$ mag 
(i.e. a factor of 101), 
$\Delta$ H $\simeq 4.27$ mag, 
and $\Delta $K$_{\rm s} \simeq 3.64$ mag.
Using in addition the error-weighted averages of the 2MASS and DENIS magnitudes 
for the unresolved object [MR81] H$\alpha$ 17, we obtain 
K$_{\rm s}$ = $13.24 \pm 0.05$ mag, H = $14.4 \pm 0.05$ mag, 
and J = $15.96 \pm 0.05$ mag for the object NE,
and K$_{\rm s}$ = $9.60 \pm 0.02$ mag, H = $10.13 \pm 0.03$ mag, 
and J = $10.96 \pm 0.01$ mag for the object SW.

\begin{figure}
\includegraphics[angle=0,width=1\hsize]{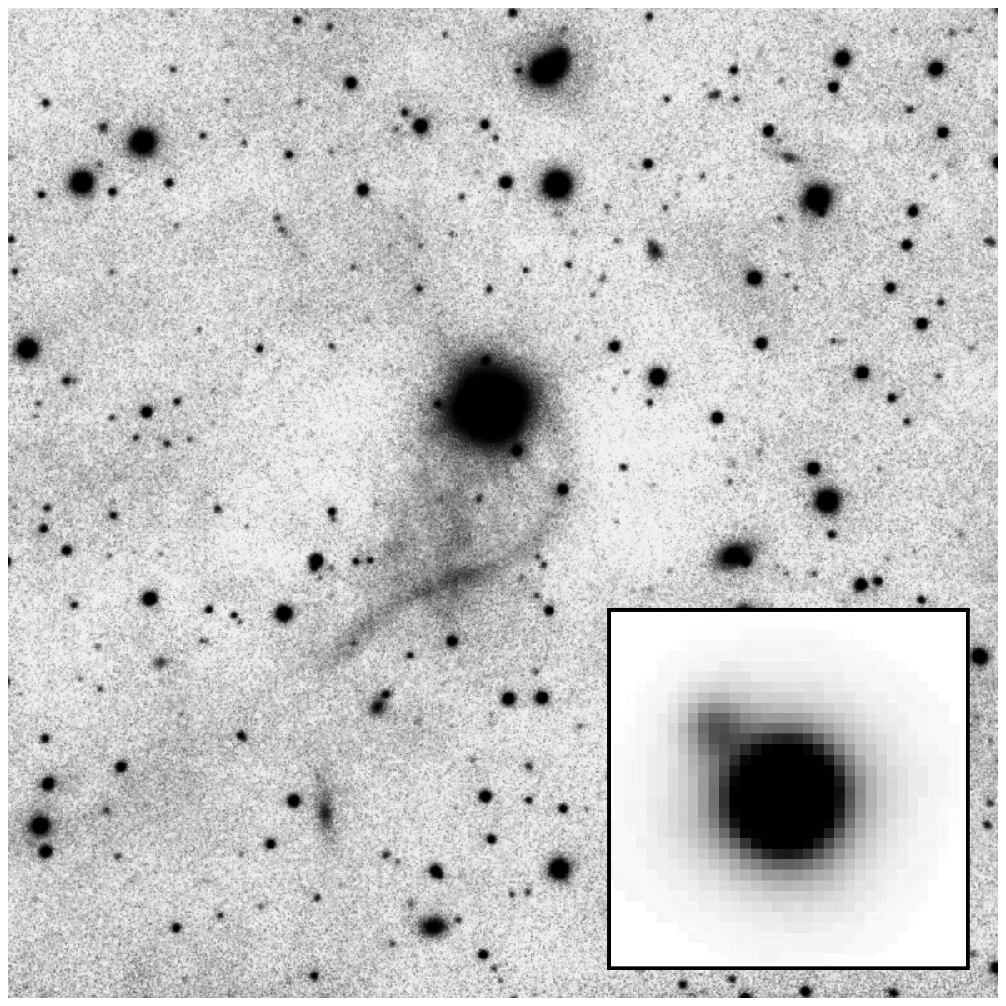}
\caption{Our VLT/ISAAC image of [MR81] H$\alpha$ 17 (central bright star)
with the filamentary structure towards the south. 
The binary is unresolved in the main image due to cuts selected to
see the filament, but seen when displayed with other cuts 
(lower right box with $5^{\prime \prime} \times 5^{\prime \prime}$ field of view).
The main image field size is $126^{\prime \prime} \times 126^{\prime \prime}$,
north is up, east is to the left.
See Fig. 14 in the online appendix for a JHK color composite; 
the image shown above is a contrast-enhanced B\&W version of the JHK color composite.}
\end{figure}

In the ISAAC JHK images, we observe a filament in addition to [MR81] H$\alpha$ 17 SW and NE
(see Fig. 6), that is most pronounced in the J and H-bands, but also visible in K$_{\rm s}$;
it is also marginally visible in the 2MASS JHK images (lower sensitivity).
Given the visual appearence, one is tempted to speculate that the filament originates
in [MR81] H$\alpha$ 17; we observe a helix- or spiral-like structure going 
first towards SW for $\sim 14^{\prime \prime}$, then appears to curve to the SE
(detected in our ISAAC images until $\sim 37^{\prime \prime}$ separation 
from [MR81] H$\alpha$ 17, or $\sim 5000$ AU at 130 pc); 
it is detected even out to $\sim 170^{\prime \prime}$ 
separation towards SE ($\sim 22360$ AU at 130 pc) in the 2MASS images.
Strong (forbidden) emission lines, as detected 
in [MR81] H$\alpha$ 17 NE, are often related to outflows, so that a physical
connection is possible, but uncertain. The curvature could be produced by 
rotational and/or orbital revolution. However, the proper motion of 
the binary is directed towards the south, 
i.e. not away from the filament in the plane of the sky,
but it is of course possible that the (as yet unknown) radial (or orbital ?) 
velocity of [MR81] H$\alpha$ 17 NE ensures that the filament trails behind.
It would be important to complete a narrow-band imaging surbey of this
feature and to monitor its possible variability.

\subsection{Spectroscopy results}

The optical spectrum was aqquired on 27 Mar 2003 with VLT/FORS1 (three exposures of
790s each at airmass 1.2 to 1.3) with the $0.7^{\prime \prime}$ slit and 
the order-sorting filter OG590 and grism 300I, i.e. covering 6000\AA~to 1.1$\mu m$ with
a dispersion of 2.5\AA~resulting in a spectral resolution of between 6.4 and 11\AA .
The faint object NE was located directly on the slit, and the brighter object SW
was also close to the slit, so that we could also extract its spectrum.
The optical spectra are shown in Figs. 7 \& 8.

In the optical spectrum, we observe remarkable emission lines
with H$\alpha$ being clearly the strongest of all. 
The remainder of the spectrum is dominated by strong emission 
of forbidden lines such as [OI], [NII], and [SII]. 
These forbidden lines are usually signs of 
shocked low-density regions of young stars
such as outflows, winds, and jets (see e.g. Cabrit et al. (1990), 
Hamann (1994), Hartigan et al. (1995), Hirth et al. (1997), 
Dougados et al. (2000), and Fern\'andez \& Comer\'on (2001)).
Outflow is often seen as both an indication and result of ongoing accretion.

The optical spectrum with these strong emission lines looks generally
similar to young, low-mass objects like HBC 617 in 
Lupus 
(Krautter et al. 1984),
LS-RCrA 1 in CrA (Fern\'andez \& Comer\'on 2001; Comer\'on \& Fern\'andez 2001),
as well as Sz 102, Sz 106, and Par-Lup3-4 in Lupus (Comer\'on et al. 2003),
e.g. our object [MR81] H$\alpha$ 17 NE has an H$\alpha$ equivalent width
of $\sim 330$ \AA~(but may be slightly saturated), while LS-RCrA 1 has 360 \AA .
Fern\'andez \& Comer\'on (2001) concluded that LS-RCrA 1 is surrounded
and extincted by an edge-on disk, which was not resolve by their imaging data.

We list all emission lines for which we care able to measure reliable equivalent widths 
in Table 3. Between 7319 and 7329 \AA , the noise is too strong, and/or 
possible lines to weak, and/or spectral resolution too poor 
to unambiguously identify feature(s) either as [CaII] line at 7324\AA~or
as a [OII] doublet at 7319 and 7329 \AA .
The CaII triplet at 8498 to 8662 \AA~is detected.

The five strongest emission lines in [MR81] H$\alpha$ 17 NE are
saturated, and their equivalent width measurements therefore represent 
uncertain limits.
There even appear 
Fake emission lines even appear at these wavelengths
in the other object [MR81] H$\alpha$ 17 SW, which was located
outside the slit. Therefore, it is also difficult to measure the
H$\alpha$ equivalent width of the object SW (whose upper limit should be
$\sim 16$ \AA ), and we are unable to classify unambiguously the SW object 
as being either a classical or weak-line T Tauri star.
For the BVRIJHK part of the spectral energy distribution,
the SW object resembles a blackbody; the IRAS 12 to 60~$\mu m$
data are offset above the blackbody (see Fig. 10), but it is unclear,
whether the far-IR emission originates in the immediate surroundings
of the star (its disk) or its more distant surroundings
(the star-forming cloud). The object [MR81] H$\alpha$ 17 NE can
clearly be classified as a classical T Tauri star because of its
strong H$\alpha$ emission and edge-on disk.

\begin{figure}
\includegraphics[angle=0,width=1\hsize]{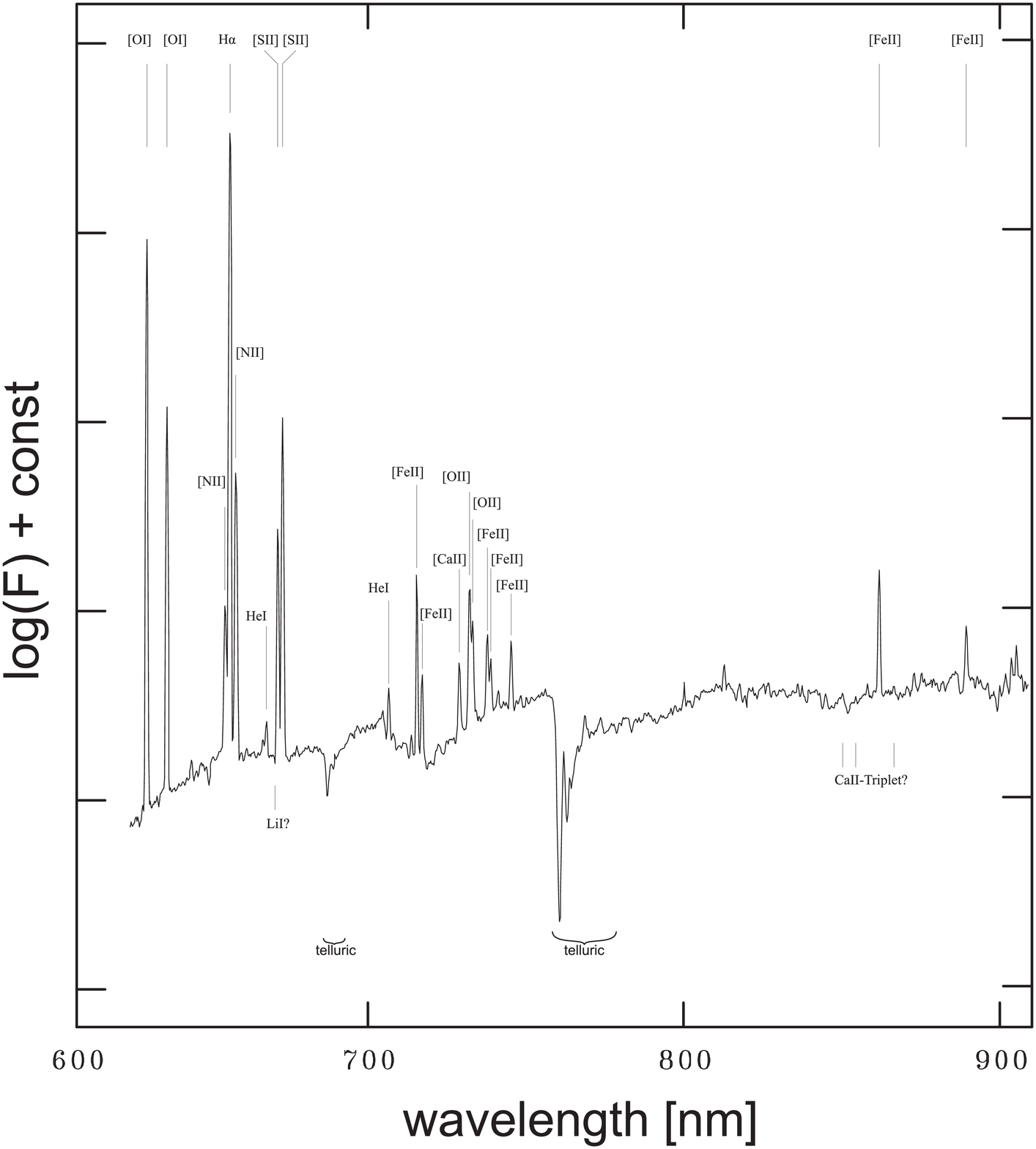}
\caption{Optical VLT/FORS1 spectrum of [MR81] H$\alpha$ 17 NE 
with normalized flux versus wavelength (in nm). 
Several strong emission lines are indicative of ongoing accretion.} 
\end{figure}

\begin{figure}
\includegraphics[angle=0,width=1\hsize]{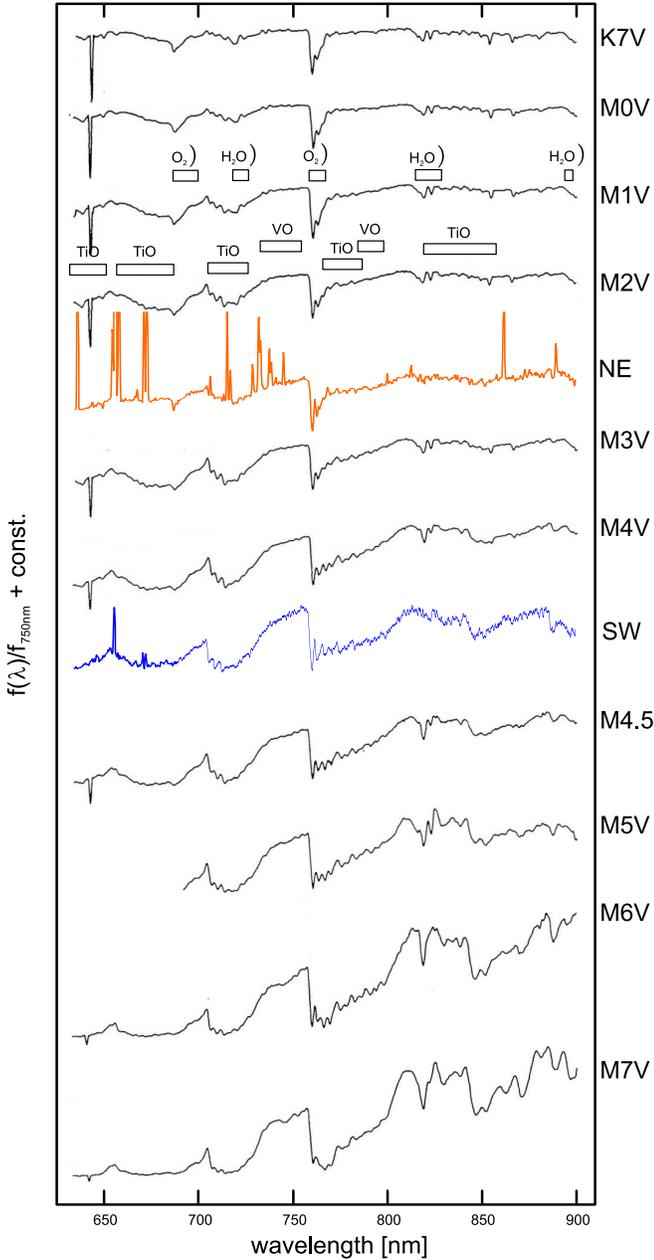}
\caption{Spectra of standard stars and [MR81] H$\alpha$ 17 NE and SW
(marked NE and SW), normalized flux (shifted in y-axes to separate
spectra) versus wavelength (in nm). We obtain spectral types to be M2e for the 
apparently fainter object NE and M4-5e for the apparently brighter 
object SW. We indicate the location of telluric water and oxygen lines 
in the top row (between M0 and M2) and the location of detected molecular 
TiO and VO bands in the 2nd row (between M1 and M2).}
\end{figure}

\begin{table}
\caption[]{Lines detected in [MR81] H$\alpha$ 17 NE} 
\begin{tabular}{lccc} \hline
Line & $\lambda_{0}$ & $\lambda$ & equ. width \\
     & [\AA ] lab    & [\AA ] obs & [\AA ] \\ \hline
$[$OI$]$   &  6300   &  6301.0 & $-205:$ \\ 
$[$OI$]$   &  6364   &  6365.0 & $-69:$ \\
$[$NII$]$  &  6548   &  6548.6 & $-11$ \\
H$\alpha$  &  6563   &  6564.1 & $-330:$ \\
$[$NII$]$  &  6583   &  6584.3 & $-36$ \\
HeI        &  6678   &  6679.7 & $-2$ \\ 
$[$SII$]$  &  6716   &  6716.7 & $-21:$ \\
$[$SII$]$  &  6730   &  6731.1 & $-47:$  \\
HeI        &  7066   &  7067.2 & $-2$  \\
$[$FeII$]$ &  7155   &  7156.3 & $-13$  \\
$[$FeII$]$ &  7452   &  7453.5 & $-3$ \\
$[$FeII$]$ &  8617   &  8618.3 &  $-9$ \\
$[$FeII$]$ &  8892   &  8893.8 &  $-2$  \\ \hline
\end{tabular}

Note: Colons indicate uncertain limits due to saturation.
\end{table}

Due to poor spectral resolution, strong (saturated) nearby [SII] emission lines,
and strong veiling in the NE object (and low S/N data of the SW object),
the Lithium 6708\AA~line is detected only marginally in both
the SW and NE components.

We determine the spectral types of both objects by comparison with
M dwarfs of known spectral types from Kirkpatrick et al. (1991) (see Fig. 8).
Surprisingly, the spectral type of the apparently fainter component (NE)
is earlier (M2e) than that of the apparently brighter component (SW),
which is M4-5e according to our spectra. These spectral types are 
consistent with the results of Patten (1998), 
who assigned M3-5e to the unresolved system.

We display the K-band spectra of both objects in Fig. 9, both of which 
exhibit the CO lines typical of late-type objects. 
The spectra of [MR81] H$\alpha$ 17 NE also
exhibit emission lines of H$_{2}$ and Br $\gamma$. 
As in LS-RCrA 1 (Fern\'andez \& Comer\'on 2001), the H$_{2}$ lines
are stronger than the Br $\gamma$ line, but in the data of our object, the
Br $\gamma$ line is clearly detected and there are more H$_{2}$
emission lines detected than in LS-RCrA 1.
These H$_{2}$ emission lines are also observed in other young
low-mass objects surrounded by dense circumstellar material
such as GY 11 (Greene \& Lada 1996; Wilking et al. 1999).
Our VLT ISAAC H-band spectra of both objects are basically feature-less
and triangular-shaped, as is typical of young M-type objects.
According to both the H- and K-band spectra, the NE component is
slightly earlier (early-M) in spectral type than the SW component
(mid-M), as also observed in the optical spectra.

\begin{figure}
\includegraphics[angle=270,width=1\hsize]{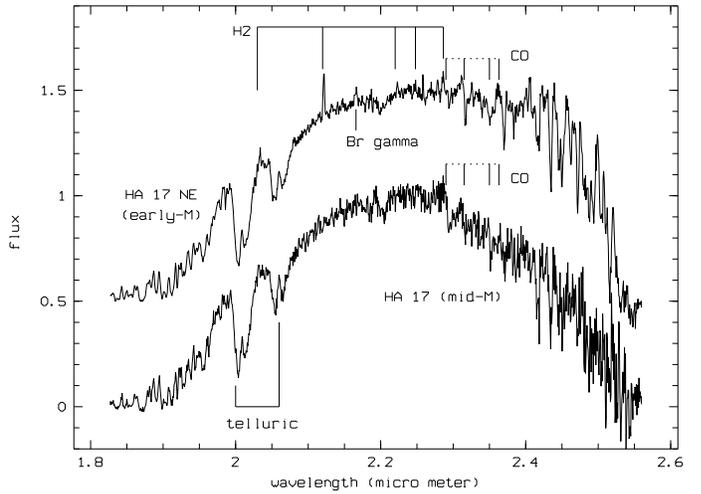}
\caption{Infrared VLT/ISAAC K-band spectra of [MR81] H$\alpha$ 17 NE (top)
and SW (bottom), flux versus wavelength (in $\mu m$), with H2 and 
Br $\gamma$ emission lines in the object NE that indicate the accretion.}
\end{figure}

Since we detect indications of youth in both objects, such as emission lines and Lithium
absorption, both objects are young and most probably CrA members; the proper motions
of both objects are also consistent with CrA membership.
If both objects are at the same distance and age, then the earlier-type component
should be more massive and intrinsically brighter than the later-type object.
Obviously, a large amount of flux from the NE component is absorbed by its edge-on disk,
and the forbidden emission lines and the strong H$\alpha$ emission 
are a clear characteristic of accretion.

\section{Discussion}

Given the spectral type and JHK colors\footnote{also optical colors BVRI 
for the unresolved object from USNO, NOMAD, GSC, and DENIS, 
which are almost identical to the colors of the 
apparently brighter object [MR81] H$\alpha$ 17 SW}, 
we can estimate the extinction towards 
the SW object and obtain 
A$_{\rm V} \simeq 1$ mag for SW;
the extinction towards the NE objects cannot be determined,
because we unablt to detect photospheric light.

\begin{figure}
\includegraphics[angle=270,width=8.5cm]{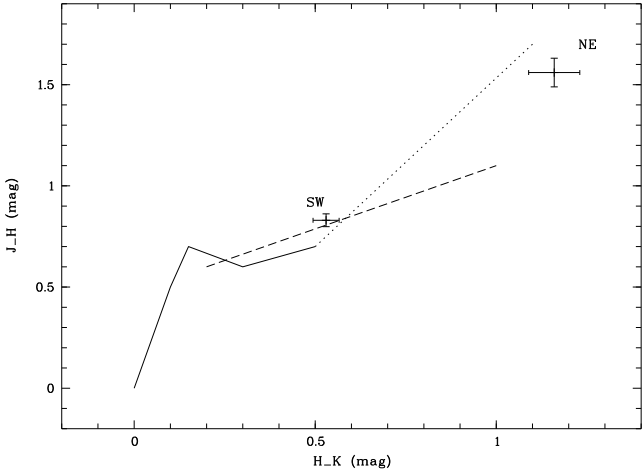}
\caption{JHK color-color diagram for [MR81] H$\alpha$ 17 SW and NE
(plotting J-H versus H-K, both in mag).
The full curved line indicates the loci of unreddened M dwarfs
down to M6 (Bessell \& Brett 1988); the dotted line is the
reddening vector for 10 mag of extinction;
and the dashed line indicates the loci of classical T Tauri stars 
(Meyer et al. 1997). Objects located to the right of the reddening vector
are surrounded by circumstellar material producing near-IR excess emission.
Our object SW is slightly extincted, but does not exhibit a near-IR excess.
The fainter object NE, however, appears strongly extincted and
also shows some near-IR excess, which is typical of T Tauri stars with disks.}
\end{figure}

In the JHK color-color diagram, both objects lie on 
or close to the reddening vector, which has its origin at a mid-M dwarf
(see Fig. 10).
The object SW is located at the intersection of the reddening vector for dwarfs
and the loci of classical T Tauri stars (Meyer et al. 1997),
i.e. there is no evidence of near-IR excess emission from the SW component.
The NE object lies 10 mag of extinction away from the 
unabsorbed mid-M dwarf and also about one tenth of a magnitude
to the right (red) of the reddening vector, which indicates
strong 
near-IR excess emission typical of circumstellar material
(and a disk).
See also Fig. 11 for the spectral energy distribution of [MR81] 
H$\alpha$ 17 SW, which again shows its spectral type to be M4-5,
but no near-IR excess emission.

\begin{figure}
\includegraphics[angle=270,width=1\hsize]{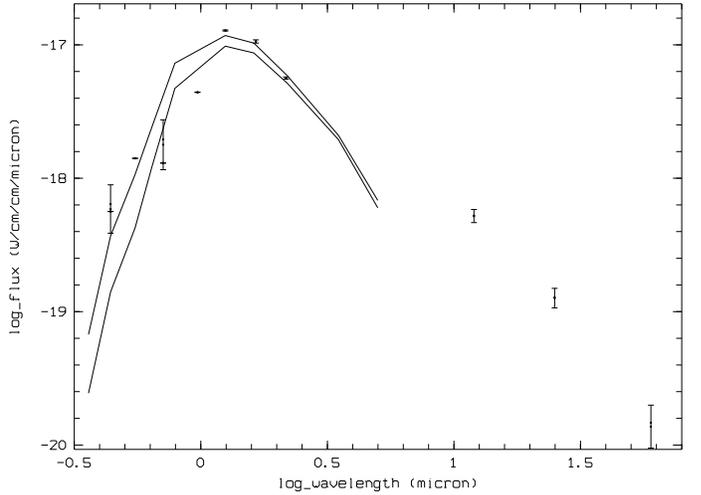}
\caption{Spectral energy distribution of [MR81] H$\alpha$ 17 SW
with data from GSC, USNO, NOMAD, and this paper
(error bars include measurement errors, when available, and variability)
as well as IRAS together with M4 and M5 standard stars. 
We plot log of flux in $W/cm^{2}/\mu $m versus log of wavelength in $\mu $m.
The brighter object SW shows no near-IR excess.
There is strong far-IR excess at the IRAS data points plotted here,
but the spatial resolution of IRAS is
not sufficient to conclude that H$\alpha$ 17 SW has IRAS excess emission;
the IRAS data plotted here are for IRAS source F19073-3704 (Moshir et al. 1989).
}
\end{figure}

We then calculate their bolometric luminosities (at 130 pc),
obtaining $\log~(L_{\rm bol}/L_{\odot}) \simeq -0.9 \pm 0.1$ 
for the apparently brighter SW (M4-5e, hence T$_{\rm eff} \simeq 3220$ K)
and $\log~(L_{\rm bol}/L_{\odot}) \simeq -1.8 \pm 0.2$ 
for the apparently fainter NE (M2e, hence T$_{\rm eff} \simeq 3560$ K),
using a temperature scale for young M-type objects that is
intermediate between dwarfs and giants from Luhman (1999).
However, since the faint NE object is detectable only in scattered light
due to the edge-on disk, we are unable to study the stellar photosphere,
so that the aforementioned luminosity for the object NE is not the 
luminosity of the star. This luminosity value appears strongly under-luminous, 
which it is typical of objects with edge-on disks,
(see e.g. Watson \& Stapelfeldt (2007)).

\begin{figure}
\includegraphics[angle=0,width=1\hsize]{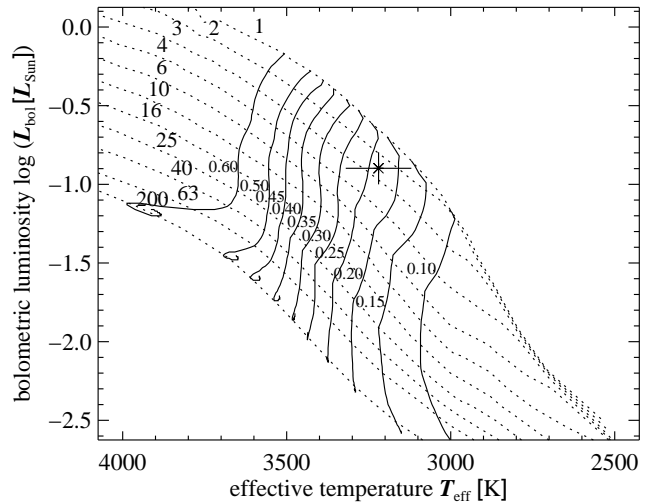}
\caption{H-R diagram with the location of [MR81] H$\alpha$ 17 SW, 
bolometric luminosity versus effective temperature,
compared to theoretical tracks (full lines with masses given
in solar masses) and isochrones (as dashed lines with age given
in Myrs) from Baraffe et al. (1998), 
version with Helium content Y=0.275, 
metallicity [M/H]=0, and mixing length $\alpha = 1.0$.
We conclude that SW has an age of 2-3 Myrs and a mass of $\sim 0.23$~M$_{\odot}$.
NE cannot be plotted here, because we see it only in reflected
light, so that we cannot measure its luminosity.
}
\end{figure}

The SW object (M4-5e) lies on the 2-3 Myr isochrone
of the Baraffe et al. (1998) models (see Fig. 12) and on the 3 Myr isochrone
on both the Burrows et al. (1997) and D'Antona \& Mazzitelli (1997) models.
According to those models, it lies between the 
(interpolated) 0.15 to 0.30~M$_{\odot}$ tracks.
Hence, we obtain an age of 2-3 Myrs and 
a mass of $0.23 \pm 0.05$~M$_{\odot}$ for this object (SW).
Given its temperature, the
NE object lies on the $\sim 0.3$ to 0.5~M$_{\odot}$ tracks,
according to Baraffe et al. (1998) and D'Antona \& Mazzitelli (1997),
almost independently of luminosity, which cannot be measured.
Hence, the apparently fainter object (early-M) is more massive than
the apparently brighter object (mid-M), 
$\sim 0.5$~M$_{\odot}$, if coeval with the SW object.

\section{Summary and conclusion}

Based on imaging and spectroscopic observations in the red optical
and near-infrared, we conclude that the object [MR81] H$\alpha$ 17 NE
is surrounded (and absorbed) by a disk seen nearly edge-on. 
The NE object appears fainter and has a spectral type M2e,
while the object SW appears brighter, but has spectral type M4-5e.
Both objects exhibit indicatons of youth such as emission lines and
common proper motion with known CrA members,
i.e. they are both low-mass T Tauri stars
(both exhibit strong H$\alpha$ emission).

The astrometric data from SHARP 2001 to NACO 2008 
indicate that the two objects separated by $\sim 1.3^{\prime \prime}$ 
exhibit common proper motion. When using only AO data from NACO, i.e. 
data of the highest quality pixel scale and resolution, and 
aquired by an identical instrument, then 
we have strong evidence for common proper motion.
Even with similar proper motion, 
they could be independent members of the young CrA star-forming region.
In both cases, they have an age of 2-3 Myrs at $\sim 130$ pc.

A more detailed investigation of the emission lines is beyond
the scope of this paper, and should be completed together with a study of 
their expected variability and higher-resolution spectra.
Our spectral resolution is insufficient to resolve the
line shape, e.g. of an inverse P Cyg profile, or to measure accurately outflow velocities.
Using data of even higher resolution and/or deeper studies in the future,
which cover a wider wavelength range, e.g. from space, one can investigate in detail the 
density distribution, scale height, and the gas and dust population, 
possibly polarization, accretion, outflow,
variability, and ongoing planet formation in the disk.

\begin{acknowledgements}
We would like to thank the ESO User Support Group and ESO service mode observers
for their support and observations. We have made use of the 2MASS, DENIS, USNO,
GSC, NOMAD, and IRAS Faint Source Catalogs as well as the Simbad, Visir, and ADS databases.
RN wishes to acknowledge general support from the German National Science Foundation
(Deutsche Forschungsgemeinschaft, DFG), grants NE 515/13-1, 13-2 and 23-1.
MAvE was supported first by a graduate scholarship of the Cusanuswerk, 
one of the national student elite programs of Germany, 
and then by an individual fellowship granted by the Funda\c{c}\~ao para a
Ci\^{e}ncia e a Tecnologia (FCT), Portugal (reference SFRH/BPD/26817/2006).
TOBS acknowledges support from Evangelisches Studienwerk e.V. Villigst,
another national student elite program in Germany.
NV acknowledges support by grants FONDECYT 1061199 and DIPUV 07/2007.
We also thank our referee, Karl Stapelfeldt, for very helpful remarks.
\end{acknowledgements}

\section{Online Appendix}

\begin{figure*}
\includegraphics[angle=0,width=1\hsize]{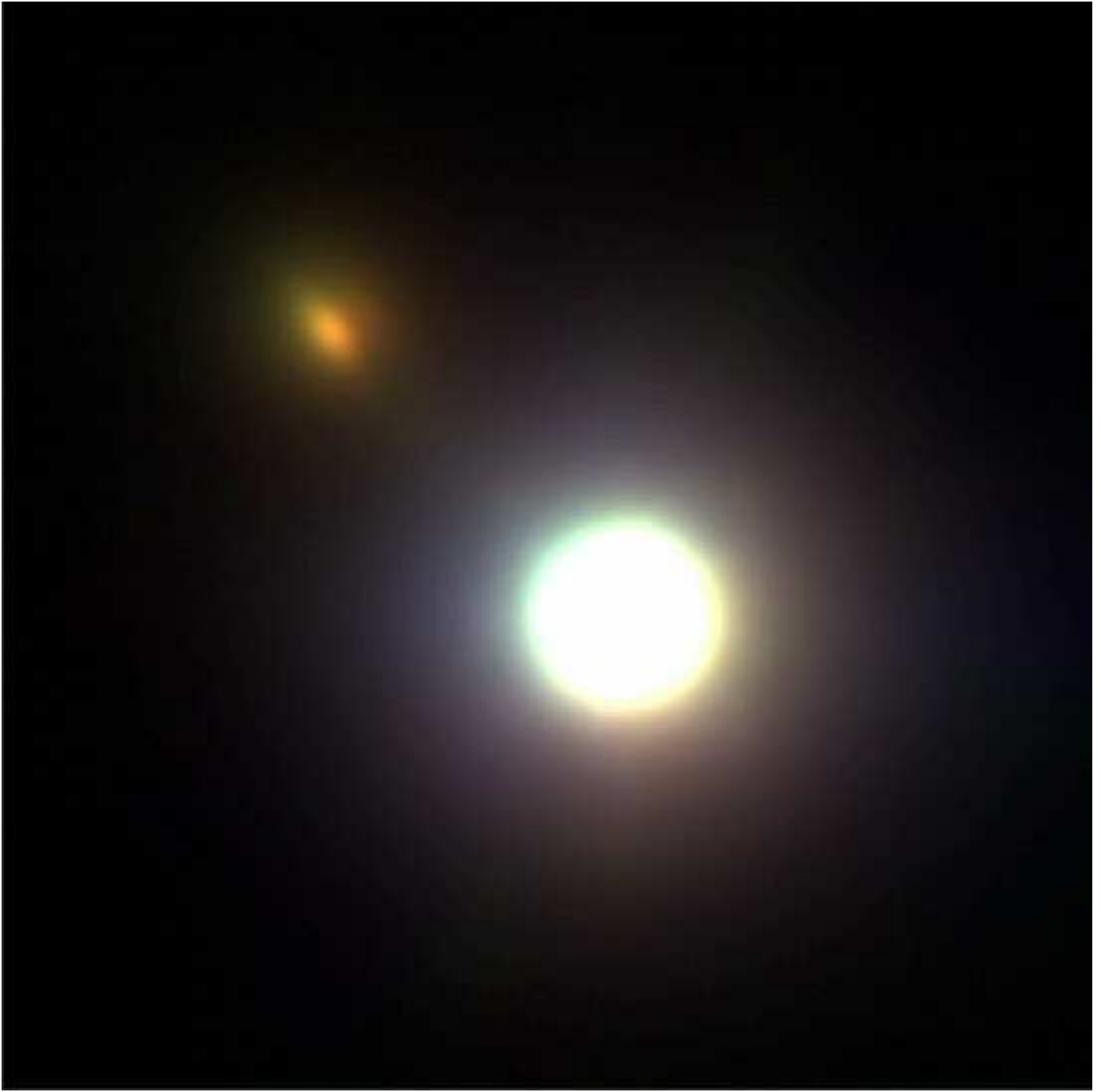}
\caption{JHK color composite of our VLT/NACO images of [MR81] H$\alpha$ 17 SW
(brighter object, K=9.60 mag) and NE (top left, K=13.24 mag, 
separated $\sim 1.3^{\prime \prime}$) using the J- and H-band images 
obtained in June 2008 and the K-band image obtained in June 2004 under
better seeing conditions (than in June 2008).
This nearly edge-on disk is seen as dark lane in front 
of the fainter object along the NE-SW direction. 
See Fig. 3 for a 3D contour plot.
Obtain tar archive from www.astro.uni-jena.de/Users/rne/cra-disk 
with a higher-resolution version of this image.}
\end{figure*}

\begin{figure*}
\includegraphics[angle=0,width=1\hsize]{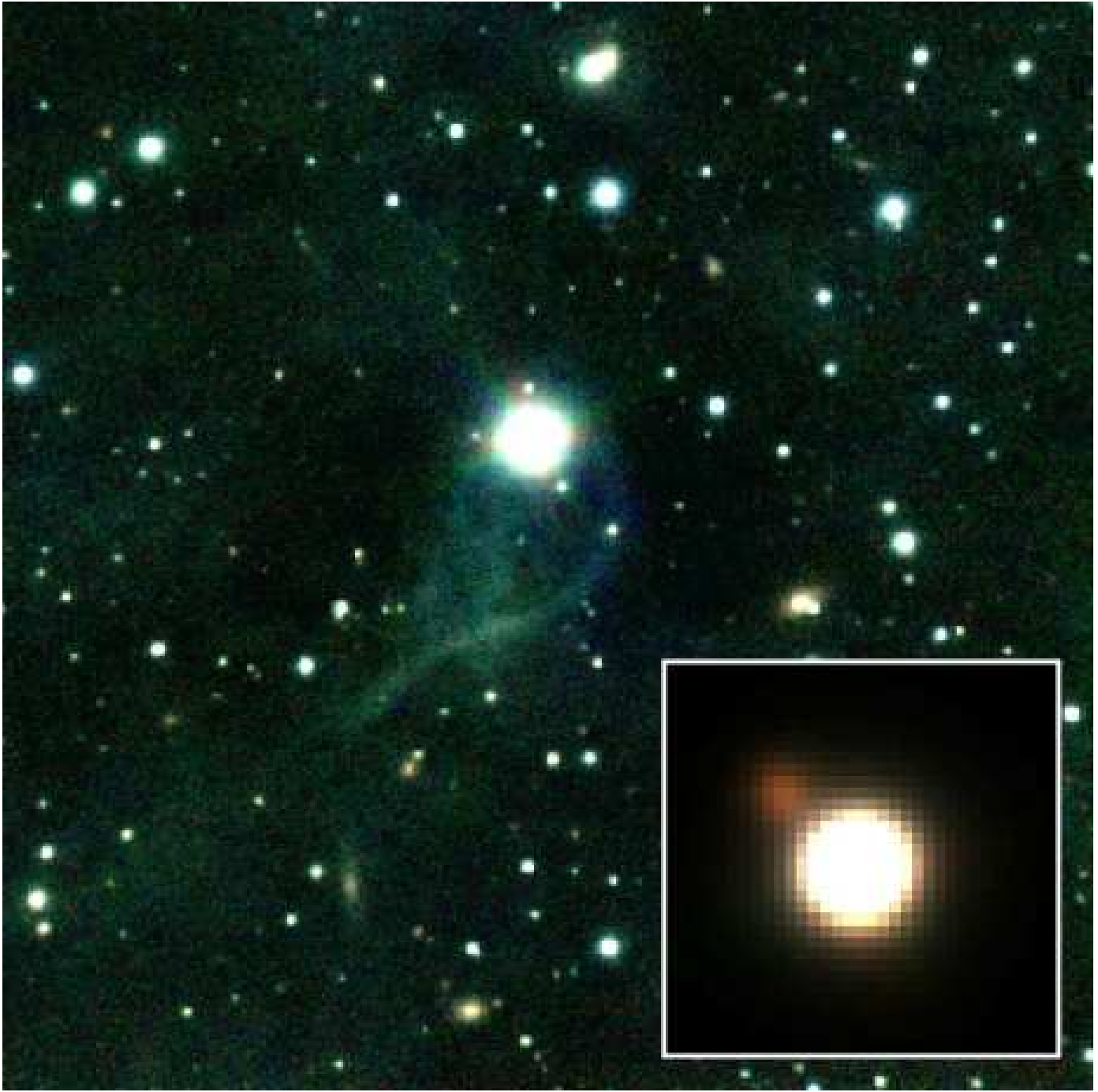}
\caption{JHK color composite of our VLT/ISAAC images 
of [MR81] H$\alpha$ 17 (central bright star)
with the filamentary structure towards the south.
The binary is unresolved in the main image due to cuts selected to 
see the filament, but seen when displayed with other cuts
(lower right box with $5^{\prime \prime} \times 5^{\prime \prime}$ field of view).
The main image field size is $126^{\prime \prime} \times 126^{\prime \prime}$,
north is up, east to the left.
Obtain tar archive from www.astro.uni-jena.de/Users/rne/cra-disk
with a higher-resolution version of this image.}
\end{figure*}

\end{document}